\documentclass[preprint,12pt,authoryear]{elsarticle}

\usepackage{amssymb}
\usepackage{amsmath}

\usepackage{threeparttable}
\usepackage{array}

\usepackage[margin=1in]{geometry}  

\usepackage{algorithm}
\usepackage{algpseudocode}
\usepackage{xcolor,soul}

\definecolor{comment}{rgb}{0.0, 0.6, 0.0} 
\definecolor{function}{rgb}{0.5, 0.0, 0.8} 
\definecolor{purple}{rgb}{0.5, 0.0, 0.8} 
\definecolor{number}{rgb}{0.3, 0.6, 1.0} 
\definecolor{parallel}{rgb}{0.6, 0.0, 0.0} 
\definecolor{input}{rgb}{1.0, 0.0, 0.5} 
\definecolor{output}{rgb}{1.0, 0.0, 0.5} 

\algdef{SE}[FOR]{ParFor}{EndParFor}[1]{\textcolor{parallel}{\textbf{parfor}}\ #1\ \textcolor{parallel}{\textbf{do}}}{\textcolor{parallel}{\textbf{end parfor}}}

\usepackage{lineno}

\usepackage[colorlinks=true,linkcolor=blue,citecolor=blue,urlcolor=blue]{hyperref}

\journal{Computer and Fluids}

\begin{document}

\begin{frontmatter}

\title{A GPU-based Compressible Combustion Solver for Applications Exhibiting Disparate Space and Time Scales}

\author[affil1]{Anthony Carreon\corref{cor1}}
\ead{acarreon@umich.edu}
\author[affil1]{Jagmohan Singh}
\author[affil1]{Shivank Sharma}
\author[affil1]{Shuzhi Zhang}
\author[affil1]{Venkat Raman}

\affiliation[affil1]{
    organization={Department of Aerospace Engineering, University of Michigan},
    city={Ann Arbor},
    state={Michigan 48105},
    country={USA}
}
\cortext[cor1]{Corresponding author}

\begin{abstract}
High-speed chemically active flows present significant computational challenges due to their disparate space and time scales, where stiff chemistry often dominates simulation time. While modern supercomputing scientific codes achieve exascale performance by leveraging graphics processing units (GPUs), existing GPU-based compressible combustion solvers face critical limitations in memory management, load balancing, and handling the highly localized nature of chemical reactions. To this end, we present a high-performance compressible reacting flow solver built on the AMReX framework and optimized for multi-GPU settings. Our approach addresses three GPU performance bottlenecks: memory access patterns through column-major storage optimization, computational workload variability via a bulk-sparse integration strategy for chemical kinetics, and multi-GPU load distribution for adaptive mesh refinement applications. The solver adapts existing matrix-based chemical kinetics formulations to multigrid contexts. Using representative combustion applications including hydrogen-air detonations and jet in supersonic crossflow configurations, we demonstrate $2-5\times$ performance improvements over initial GPU implementations with near-ideal weak scaling across $1-96$ NVIDIA H100 GPUs. Roofline analysis reveals substantial improvements in arithmetic intensity for both convection ($\sim 10 \times$) and chemistry ($\sim 4 \times$) routines, confirming efficient utilization of GPU memory bandwidth and computational resources.
\end{abstract}

\begin{graphicalabstract}
\includegraphics{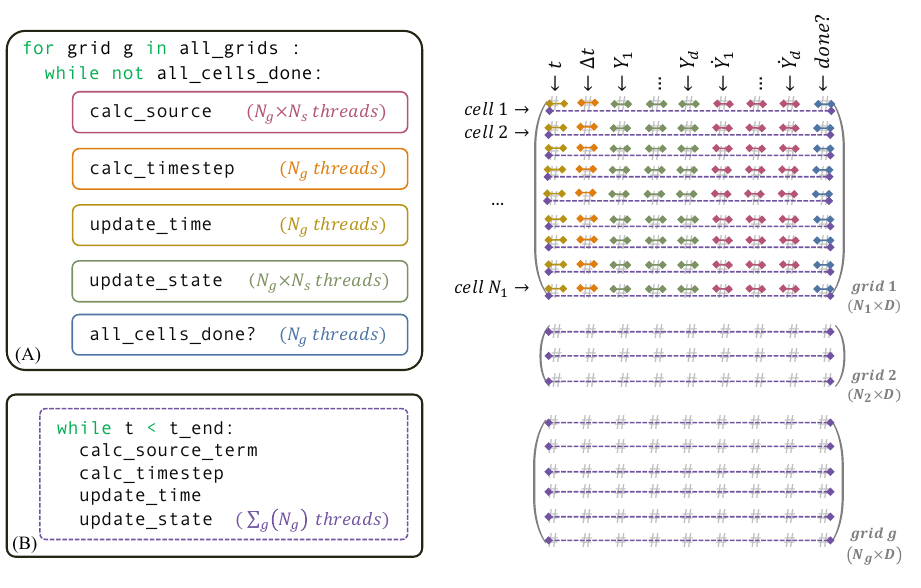}
\end{graphicalabstract}

\begin{highlights}
    \item A GPU-optimized compressible combustion solver built on AMReX achieves $2-5 \times$ speedup over initial implementations
    \item Novel bulk-sparse chemical kinetics integration strategy plus multigrid kernel fusion resulted in up to a $6 \times$ speedup for the chemistry routine.
    \item Near-ideal weak scaling demonstrated across $1-96$ NVIDIA H100 GPUs with substantial arithmetic intensity improvements ($\sim 10 \times$ for convection and $\sim 4 \times$ for chemistry)
\end{highlights}


\end{frontmatter}

\section{Introduction}

Computational modeling has become indispensable for designing complex reacting flow systems, particularly where practical experiments are prohibitively expensive or infeasible \citep{raman2019emerging}. However, simulating such flows may also become computationally intractable due to disparate space and time scales, primarily driven by stiff chemistry at each computational cell \citep{raman2023nonidealities}. Chemical kinetics routines often dominate the overall simulation time, accounting for over 90\% of computational expense in detailed chemistry calculations \citep{shi2012accelerating, uranakara2023accelerating}. Recent works have demonstrated substantial acceleration for chemical kinetics on graphics processing units (GPUs), with \cite{barwey2021neural} introducing a matrix-based formulation inspired by neural network architectures. This approach was successfully adopted by \cite{uranakara2023accelerating}, achieving an overall speedup of up to 7X.

Other studies have explored GPU usage or adaptive mesh refinement (AMR) techniques for accelerating CFD codes \citep{sharma2024amrex, bernardini2021streams, lai2019multi, barwey2021neural, bielawski2023highly, romero2020zefr, witherden2015heterogeneous}. AMR is an approach that helps address the disparate space and time scales of compressible reacting flow simulations on large, yet fine, meshes by resolving only spatio-temporal regions that need higher resolution. For instance, the AMReX framework \citep{zhang2019amrex} — a structured-grid AMR coding framework — provides a hardware-agnostic solution for CFD code development, easing code portability to a diverse set of exascale architectures. AMReX-based solvers such as PeleC \citep{henry2023pelec} have demonstrated remarkable scalability. However, achieving optimal GPU performance demands algorithmic restructuring and hardware-specific tuning beyond CPU-GPU porting strategies provided by AMReX.

GPU-based optimizations for high-speed reacting flow solvers remain under-explored, particularly in terms of combining chemical kinetics with mult-grid AMR algorithms. Key challenges include: (1) memory access pattern optimization for hierarchical grid structures, (2) managing computational workload variability arising from the highly localized and temporally variable nature of chemical reactions in AMR contexts, (3) kernel-level optimizations that improve computational intensity, and (4) multi-kernel control flow that reduces kernel launch overhead, warp divergence, and global memory access. \cite{wang2025efficient} addressed points (1) and (3) by focusing on low-storage integration schemes that reduce register usage to increase GPU occupancy; however, system-level performance bottlenecks unique to multi-grid AMR contexts remain unaddressed.

To this end, we present a GPU-optimized AMReX solver for compressible reactive flows. Our implementation alleviates computational bottlenecks associated with chemistry via algorithmic restructurings and GPU-specific tuning, setting the stage for future scientific supercomputing endeavors. The remainder of this paper reads as follows: Section~\ref{sec_background} presents background material, including the governing equations, GPU concepts, and the AMReX framework. Sections~\ref{sec_gpu_implementation}~and~\ref{sec_setup_for_performance_analysis} present our GPU implementation strategies and the setup for our performance analysis. Section~\ref{sec_results} demonstrates the solver's performance improvements across multiple GPUs through simulations of 2D detonations and a jet-in-crossflow. We also include a roofline analysis. Finally, Section~\ref{sec_conclusion} summarizes our findings and outlines directions for future work.

\section{Background}\label{sec_background}

\subsection{Numerics and Governing Equations}\label{sec_governing_equations}

Using AMReX, we developed a high-speed, reacting flow solver written in \texttt{C++}. The interested reader may find details of the numerical methods used by the solver in \cite{sharma2024amrex}. The computational model uses a 2nd-order finite volume method to solve the compressible Navier-Stokes equations with chemical production and transport:

\begin{equation}\label{eqn_continuity}
    \frac{\partial \rho}{\partial t}+\frac{\partial}{\partial x_i}\left(\rho u_i\right)=0 
\end{equation}
\begin{equation}\label{eqn_mom}
    \frac{\partial}{\partial t}\left(\rho u_i\right)+\frac{\partial}{\partial x_j}\left(\rho u_i u_j\right) =\frac{\partial \tau_{i j}}{\partial x_j}-\frac{\partial p}{\partial x_i}
\end{equation}
\begin{equation}\label{eqn_engy}
    \frac{\partial}{\partial t}(\rho E)+\frac{\partial}{\partial x_j}\left(\rho u_j E+u_j p\right) =\frac{\partial}{\partial x_j}\left(u_j \tau_{i j}\right)+\frac{\partial}{\partial x_j}\left(\alpha \frac{\partial T}{\partial x_j}\right)
\end{equation}
\begin{equation}\label{eqn_sps_transport}
    \frac{\partial}{\partial t}\left(\rho Y_k\right)+\frac{\partial}{\partial x_j}\left(\rho u_j Y_k\right) =\frac{\partial}{\partial x_j}\left(\rho D \frac{\partial Y_k}{\partial x_j}\right)+\rho \dot{\omega}_k,
\end{equation}

where $\rho$, $u_i$, $p$, $E$, and $T$ represent the mass density, $i$-th velocity component, pressure, specific total energy, and temperature, respectively. $\alpha$ is the thermal diffusivity. $\tau_{i j}$ represents the stress tensor, where a Newtonian fluid assumption relates the bulk and dynamic viscosity. In Equation~\ref{eqn_sps_transport}, $Y_k$ represents the mass fraction of species $k$, $D$ is the species mixture diffusivity, and $\dot{\omega}_k$ is the production rate of species $k$.

Convective fluxes computations use the Harten-Lax-van Leer-Contact (HLLC) scheme \citep{toro1992weighted}, while diffuse fluxes use a spatially centered 2nd-order scheme. Time stepping employs a 2nd-order accurate, total-variation-diminishing Runge-Kutta method \citep{gottlieb1998total}. During a single time step, Strang operator splitting separates the flow system from the chemical system \citep{macnamara2016operator}. Since the time scales for chemical kinetics (on the order of $10^{-14}$~s) are much smaller than the flow time scales (on the order of $10^{-9}$~s), we assume the flow field remains static during local time integration of finite-rate chemistry. The in-house solver employs an adapted version of the GPU-accelerated kinetics solver developed by \cite{barwey2021neural}. At a given time step determined by the CFL stability criteria, $\Delta t_\text{CFL}$, each cell in the domain simulates a $0$D reactor, assuming constant volume and the ideal gas law. For a problem consisting of $N_s$ species and $N_c$ cells, the following equations are solved:

\begin{equation}\label{eqn_mass_fracs}
\frac{dY_{i,k}}{d t}= \frac{W_k~\Omega_{i,k}}{\rho_i}
\end{equation}

\begin{equation}\label{eqn_temp_ode}
\frac{d T_i}{d t}=-\sum_{k=1}^{N_s} \frac{Y_{i, k}}{c_{v, k}\left(Y_{i, k}, T_i\right)} \sum_{k=1}^{N_s} \epsilon_k\left(T_i\right) \Omega_{i, k}
\end{equation}

\begin{equation*}
0 < t_i <  \Delta t_{CFL}
\end{equation*}

\begin{equation*}
i=1,2,\dots,N_c-1,N_c~~~~~k=1,2,\dots,N_s-1,N_s
\end{equation*}

$i$ is the cell index while $k$ is the species index. $Y$, $\Omega$, $c_v$, $\epsilon$, $W$, $t$, $\rho$, and $T$ respectively represent the mass fractions, molar production rates, mass-based constant volume specific heat, molar internal energy as a function of temperature, and molecular weight, local time, mass density, and temperature. The kinetics routine numerically integrates Equation~\ref{eqn_mass_fracs} using an explicit 1st-order adaptive time-step scheme. The local time-step, $\Delta t_i \ll \Delta t_\text{CFL}$, is calculated such that the species mass fractions do not change by more than a set percentage (ranging from $1-5\%$) of their current value to maintain numerical stability and accuracy. The temperature calculation employs a Newton-Raphson iterative procedure, assuming a constant local internal energy and density.

\subsection{Graphics Processing Units and Roofline Models}\label{sec_overview_gpu_roofline}

Graphics processing units (GPUs) can significantly accelerate CFD calculations because many underlying algorithms can be cast into a single-instruction, multiple-data (SIMD) pattern, which GPUs are designed to exploit~\citep{flynn1972some}. While CPUs excel at complex control flow and low-latency operations, GPUs contain hundreds of processing cores termed \textit{streaming multiprocessors (SMs)} designed to execute identical operations on different data elements simultaneously. This hardware design makes them well suited for grid- and particle-based simulations, where the same numerical operations are applied independently across millions of computational elements with minimal interdependence. However, effectively utilizing GPUs requires addressing fundamental differences from CPU programming. A kernel is a function that runs many times in parallel on a GPU. The kernel executes work across many parallel workers, termed \textit{threads}. The threads work in unison within groups called \textit{warps} (32 threads per warp on NVIDIA hardware) that must execute identical instructions in lockstep. When threads within a warp take different execution paths -- a condition known as thread divergence -- performance degrades significantly. Additionally, GPUs achieve peak performance only when memory accesses follow coalesced patterns, where consecutive threads access consecutive memory locations \citep{cuda2022programming}.

In modern GPUs, performance is predominantly limited by memory bandwidth rather than arithmetic throughput \citep{nickolls2010gpu}. For instance, the NVIDIA H100 GPU \citep{h100specs} used in this study achieves peak double-precision performance of 34 TFLOPs/s but has global memory bandwidth of only 3.35 TB/s. This imbalance means that most computational kernels become memory-bound, where the costs of data movement dominate the costs of arithmetic operations. Common optimization strategies include minimizing CPU-GPU data transfers, utilizing on-chip shared memory for data reuse, ensuring coalesced memory access patterns, and maximizing arithmetic intensity—the ratio of floating-point operations to bytes transferred \citep{Schroeder2011}.

The roofline model provides an intuitive framework for understanding and optimizing GPU performance \citep{williams2009roofline}. This visual performance model plots achievable computational throughput (FLOPs/s) against arithmetic intensity (FLOPs/byte), creating a "roof" that bounds kernel performance based on hardware limitations. The model consists of two key components: a diagonal line representing memory bandwidth limitations and a horizontal line representing peak computational throughput. The intersection of these lines, known as the ridge point, indicates the minimum arithmetic intensity required to achieve compute-bound performance. Figure~\ref{fig_rooflines_all} presents an example roofline chart, where different markers represent various computational kernels. Kernels appearing below the diagonal line are memory-bound and require optimization strategies focused on data movement and memory access patterns. Kernels appearing below the horizontal line but to the right of the ridge point are compute-bound and benefit from algorithmic optimizations that reduce arithmetic complexity.

Modern roofline analysis tools, such as NVIDIA Nsight Compute \citep{NVIDIANsightCompute}, automate the collection of performance metrics necessary for roofline analysis. These tools measure kernel execution time, floating-point operations, and memory transactions to calculate both arithmetic intensity and sustained performance. The resulting roofline charts provide immediate visual feedback on performance bottlenecks and guide optimization efforts by clearly indicating whether kernel performance is limited by computational throughput, or other architectural constraints.

\subsection{Overview of AMReX}\label{sec_overview_of_amrex}

This section provides a brief overview of the AMReX framework, focusing on the key concepts and features most relevant to our GPU-accelerated combustion solver. For comprehensive documentation of AMReX's full capabilities, readers are referred to \cite{zhang2019amrex} and the official documentation \citep{amrexdocs}.

Adaptive Mesh Refinement Exascale (AMReX) is a software framework developed under the U.S. Department of Energy's Exascale Computing Project. It provides the infrastructure for block-structured adaptive mesh refinement (AMR) applications \citep{zhang2019amrex}. For multiscale problems, AMR enables dynamic grid refinement to achieve higher resolution in regions of interest (e.g., rapid variations in pressure, temperature, or chemical composition) while maintaining coarser grids elsewhere. This approach helps address the disparate space and time scales that make compressible reacting flow simulations computationally prohibitive on large, yet fine, meshes. 

The AMReX framework organizes computational cells into rectangular non-overlapping grids at different AMR levels. The coarsest level -- level 0 or the base level -- contains the base coarse grids that span the entire physical domain, while higher levels contain dynamically generated finer grids that cover only regions meeting user-defined refinement criteria. To ensure efficient multigrid performance and optimal load balancing, AMReX enforces a \textit{blocking factor} constraint, requiring all grid dimensions to be divisible by a specified power-of-two value (e.g., $2^k$ where $k \geq 0$). This constraint enables efficient grid refinement algorithms and facilitates regular memory access patterns, which are essential for achieving optimal GPU performance. From a programming standpoint, a \texttt{Fortran Array Box} (\texttt{FAB}) object efficiently stores the data and attributes for each grid, and \texttt{FAB}s at a given AMR level are collectively stored in a \texttt{multiFAB} object \citep{amrexdocs}. \texttt{multiFAB} provides a high-level interface for computational scientists to perform efficient parallel operations, load balancing, and communication. Unlike unstructured adaptive methods, this block-structured approach preserves regular data access patterns, which are essential for achieving optimal GPU performance.

AMReX also provides a framework for temporal subcycling, where finer AMR levels take multiple time steps proportional to their refinement factor, while coarser levels advance with larger time steps. For instance, level $l$ with cells $r$ times finer than those at level $l-1$ requires $r$ times as many time steps as level $l-1$.  In our solver, this capability is combined with a separate local chemical subcycling strategy (Section~\ref{sec_governing_equations}), where $\Delta t_i \ll \Delta t_\text{CFL}$, to efficiently integrate stiff reaction source terms. Together, these approaches enable stable and efficient simulations by balancing the demands of spatial refinement with large disparity between chemical and flow time scales.

\begin{figure}[h]
    \centering
    \includegraphics[width=\textwidth]{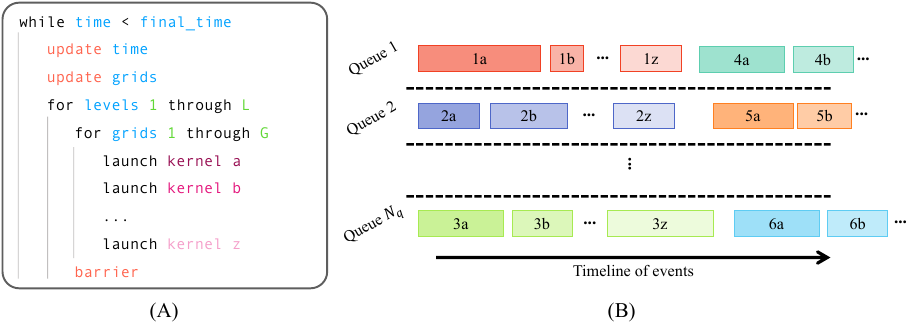}
    \caption{AMReX's GPU strategy.}
    \label{fig_amrex_gpu_strategy}
\end{figure}

For GPU computing, AMReX provides several capabilities that directly benefit our optimization efforts. The framework manages memory through pre-allocated arenas that minimize expensive allocation and deallocation operations during simulation. By default, we ensure all grid data remains in GPU memory until explicitly needed by the CPU (e.g., performing I/O operations or transferring ghost cell data between grids residing on different GPUs). AMReX also handles the complexity of GPU kernel launches through its parallel iterators. For instance, Figure~\ref{fig_amrex_gpu_strategy} shows pseudo-code for the AMReX time stepping loop across levels and grids. The \texttt{multiFAB} iterator, referred to as an \texttt{MFIter}, is used to iterate through each \texttt{FAB} and launch a sequence of kernels that operate on the underlying data. Because the \texttt{MFIter} launches the kernel sequence in one of $N_q$ CUDA streams per \texttt{FAB}, the \texttt{MFIter} loop can schedule multiple kernel sequences to run concurrently, as seen in Figure~\ref{fig_amrex_gpu_strategy}. This concurrency is beneficial in cases where individual grids may not be large enough (see \ref{apx_grid_behavior} for an example) for their kernel sequences to utilize the GPU fully, but multiple concurrent kernel sequences can. AMReX also features multigrid kernel launch functionality, which enables the launch of a single kernel for all grids at a given AMR level. While intra-grid cell data resides in contiguous memory, inter-grid data is stored in disjoint memory spaces. This non-adjacency in memory presents challenges for memory-based optimizations when kernels must operate across multiple grids simultaneously (see Sections~\ref{sec_memory_optimizations} and \ref{sec_chemistry_algorithm_optimization} for details).

Load balancing across multiple GPUs is a key challenge in multigrid solvers. In CPU settings, users can calculate computational costs for each MultiFab and provide these to AMReX's load balancing algorithms, which then redistribute the grids across the available MPI ranks to achieve balanced workloads \cite{amrexdocs}. However, this strategy becomes limited on GPUs due to the typical one-rank-per-GPU deployment, which drastically reduces the number of redistribution targets compared to CPU runs that may use dozens of ranks per node. Additionally, GPU performance is nonlinear with respect to the number of grids and cells processed, making cost estimation more complex. Load balancing becomes particularly important in scientific codes where physical features -- and thus computational costs -- can vary dramatically across the domain.

\section{Our GPU Implementation Strategy}\label{sec_gpu_implementation}

In this section, we detail our GPU implementation strategies, which resulted in significant performance improvements tailored for the H100 GPU. Broadly, our optimization efforts focused on efficient memory usage and access patterns, workload balancing, and chemistry-specific operations. We derived these optimization pathways from an extensive profiling and analysis of the solver's performance across various applications, three of which are referenced here (see Table~\ref{tab_case_params}). Note that we did not select the optimization strategies a priori or based on a single assessment of the solver's performance. Instead, after several analyze-then-improve iterations, we identified the most impactful optimizations.

\subsection{Memory-Based Optimizations}\label{sec_memory_optimizations}

AMReX already provides several GPU-based strategies for memory management and access. For instance, the framework manages memory through pre-allocated \textit{memory arenas} that reuse contiguous memory chunks to eliminate allocation overhead~\citep{amrexdocs}. By default, \texttt{amrex.the\_arena\_is\_managed} is set to $0$, allocating device-only memory. Setting it to \texttt{1} enables managed memory accessible from both the CPU and GPU, facilitating easier development but potentially incurring performance penalties due to page faults. To address the penalties, we disabled managed memory and use explicit memory management via \texttt{The\_Async\_Arena()} for temporary allocations while leveraging AMReX's arena system for memory pools. We also determined the maximum problem size for a single H100 based on the total device memory requirements for various variables. The variables include, but are not limited to, thermo-fluid variables, chemistry variables, temporary variables for flux computation, and variables for ghost cells.

By default, AMReX uses column-major storage, where all grid data for a given component is contiguous in memory. This format has an impact on GPU performance, which is dictated by memory access patterns, memory coalescence, and caching effects. In the context of our compressible combustion solver, each GPU thread operates on and accesses data for (a) all components of one cell or (b) one component of one cell. The advantage of column-major storage is that consecutive threads access consecutive cell data for a given component, which improves memory coalescence and reduces the number of global memory accesses. However, when a thread requires access to multiple components of a cell, memory coalescence may not occur, increasing the number of global memory accesses. The net benefits of column major formatting still outweigh the net benefits of row major formatting. Take, for example, Algorithm~\ref{alg_compute_internal_energy}, which computes the internal energy of a cell and is representative of many convection and chemistry routines in the solver. When one thread processes $N_v$ variables of cell $i$, column-major storage requires $\lceil N_c/32 \rceil \times N_v$ memory transactions compared to $N_c \times \lceil N_v/32 \rceil$ for row-major storage. $N_c$ is the number of cells in the grid. Because a memory request typically accesses 32 contiguous elements of data in one transaction (one for each thread in a CUDA warp), the number of memory transactions is reduced by a factor of approximately 32. For typical combustion problems where $N_v \approx 50$ and for small grids (worst-case scenario) where $N_c \approx 1024$, column-major requires $32 \times 50 = 1600$ memory transactions versus $1024 \times 2 = 2048$ for row-major, representing a 28\% difference. In Algorithm~\ref{alg_compute_internal_energy} with $N_v=6$, column-major formatting requires $\sim 5\times$ fewer transactions. As $N_c$ increases and $N_v$ decreases, this advantage becomes more pronounced.

\begin{algorithm}[H]
    \caption{GPU Kernel for Internal Energy Computation}
    \label{alg_compute_internal_energy}
    \begin{algorithmic}[1]
    \Require grid data array, $N_c$ \textcolor{comment}{// $N_c=$ total number of cells in a given grid}
    \State tid $\leftarrow$ blockIdx.x $\times$ blockDim.x + threadIdx.x \textcolor{comment}{// Calculate global thread index}
    \State
    \If{tid $< N_c$}
        \State \textcolor{comment}{// LOAD CONSERVED VARIABLES (column-major memory access)}
        \State $\rho \leftarrow$ \textcolor{purple}{data}[tid + \textcolor{number}{0} $\times N_c$] \textcolor{comment}{// Load density (column 0)}
        \State $\rho u_x \leftarrow$ \textcolor{purple}{data}[tid + \textcolor{number}{1} $\times N_c$] \textcolor{comment}{// Load momentum components (columns 1-3)}
        \State $\rho u_y \leftarrow$ \textcolor{purple}{data}[tid + \textcolor{number}{2} $\times N_c$]
        \State $\rho u_z \leftarrow$ \textcolor{purple}{data}[tid + \textcolor{number}{3} $\times N_c$]
        \State $\rho E \leftarrow$ \textcolor{purple}{data}[tid + \textcolor{number}{4} $\times N_c$] \textcolor{comment}{// Load total energy (column 4)}
        \State
        \State \textcolor{comment}{// COMPUTE KINETIC AND INTERNAL ENERGY}
        \State $u_x \leftarrow \rho u_x / \rho$ \textcolor{comment}{// Extract velocity components}
        \State $u_y \leftarrow \rho u_y / \rho$
        \State $u_z \leftarrow \rho u_z / \rho$
        \State $KE \leftarrow \textcolor{number}{0.5} \times (u_x^2 + u_y^2 + u_z^2)$ \textcolor{comment}{// Kinetic energy per unit mass}
        \State $U \leftarrow (\rho E / \rho) - KE$ \textcolor{comment}{// Internal energy per unit mass}
        \State
        \State \textcolor{comment}{// STORE RESULT}
        \State \textcolor{purple}{data}[tid + \textcolor{number}{5} $\times N_c$] $\leftarrow U$ \textcolor{comment}{// Store internal energy (column 5)}
    \EndIf
    \end{algorithmic}
\end{algorithm}

\subsection{Chemical Kinetics Solver}\label{sec_chemistry_algorithm_optimization}

The bulk of the optimization effort focused on the chemical kinetics routines. At a high level, the kinetics solver is a first-order time integrator of $N_s$-dimensional dynamical systems for $N_c$ cells all reaching the same final solution time, $t_{\text{final}}$, as shown in Algorithm~\ref{alg_naive_kinetics_solver}. $K_{\text{max}}$ is the maximum number of integration steps to safeguard against unlikely cases of non-converging chemistry. \texttt{parfor} indicates the launch of a GPU kernel. Each GPU thread is responsible for integrating the dynamical system of one cell, resulting in $N_c$ threads at kernel launch. While this algorithm can be implemented as is, it suffers from several performance deficiencies:

\begin{itemize}
    \item \textbf{Warp Divergence:} While all cells start at $t=0$ and end at the same $t_\text{final}$, the step sizes (and consequently the number of time steps needed for completion) may vary significantly across cells. This variability arises from variable species production rates, driven by reaction speeds that, in turn, are influenced by the variability in the thermochemical composition of the cells. Consequently, one may observe that some cells finish time integration in a few steps, while other cells require hundreds to thousands of steps (see \ref{apx_active_cell_count_behavior}). This variability is problematic in GPU settings as the cells that quickly reach $t_\text{final}$ may not execute the body of the \texttt{while} loop, leading to warp divergence and significant performance losses.
    \item \textbf{Inefficient GPU Utilization:} As some cells reach $t_{\text{final}}$, their corresponding GPU threads remain occupied until all cells have finished, resulting in wasteful GPU utilization and slower overall performance for high cell counts (see \ref{apx_performance_vs_active_cells})
    \item \textbf{Kernel Launch Costs:} The kernel would launch once for every grid in the domain, potentially taxing the kinetics module with significant kernel launch costs \citep{myers2024amrex}, especially for a large grid count (see \ref{apx_grid_behavior}). These costs could further increase if one attempts to mitigate warp divergence and GPU utilization inefficiencies by launching GPU kernels every time step for only incomplete cells. As such, there is a tradeoff between (i) warp divergence and inefficient GPU utilization, and (ii) kernel launch overhead, which we describe and address below.
\end{itemize}

To alleviate these deficiencies, we devised Algorithm~\ref{alg_optimized_integration}, which consists of three sections: (1) counting cells requiring reaction, (2) bulk integration on all cells, and (3) sparse integration on remaining cells. The \texttt{multiStepIntegration} function contains the source term calculations and integration steps for one cell, equivalent to the main integration loop in Algorithm~\ref{alg_naive_kinetics_solver}. $N_\text{active}$ is the number of active cells, defined as the cells that have not completed time integration. Performance results are shown and discussed in Section~\ref{sec_results}.

We address the tradeoff between (A) warp divergence and inefficient GPU utilization, and (B) kernel launch overhead by strategically selecting $K_{\text{max}}$ in the bulk integration section. $K_{\text{max}}$ can be interpreted as the maximum number of integration steps until the active cell count is updated again and the time integration kernel is re-launched. From timing experiments, $K_{\text{max}}=5$ provided a balance between kernel launch overhead costs at low $K_{\text{max}}$ values and warp divergence along with inefficient GPU utilization at high $K_{\text{max}}$ values (see Section~\ref{apx_active_cell_count_behavior}). In sparse integration, $K_{\text{max}}$ is set to $10^5$ iterations, which is sufficient for all cells to complete (see Figure~\ref{fig_reaction_clusters}).

The transition from bulk to sparse integration occurs when $N_{\text{active}}$ falls below a user-defined threshold, $N_{\text{active}}^*$, which we manually tuned to $10^4$ in our implementation (see~\ref{apx_performance_vs_active_cells}). The sparse integration section addresses the problem of wasteful GPU utilization by continuing integration on only the remaining active cells.  Finally, the algorithm restores the data for all completed cells to their original memory locations. In multigrid settings, memory copies may become expensive when working with hundreds or thousands of grids. To circumvent this, we instead utilize an index map, which indicates to each GPU thread where to locate the data for the cell it will integrate. A potential drawback to this approach is the indirect memory access, which decreases memory coalescence and increases the number of memory transactions. However, this penalty is outweighed by the elimination of expensive memory copy operations and the substantial reduction in kernel launch overhead achieved through our multigrid kernel fusion strategy, as demonstrated in Figure~\ref{fig_runtime_vs_n_grids}.

\subsubsection{cuBLAS Performance in Multi-Grid Solvers}\label{sec_cublas_par_dims}

\begin{figure}
    \centering
    \includegraphics[width=\linewidth]{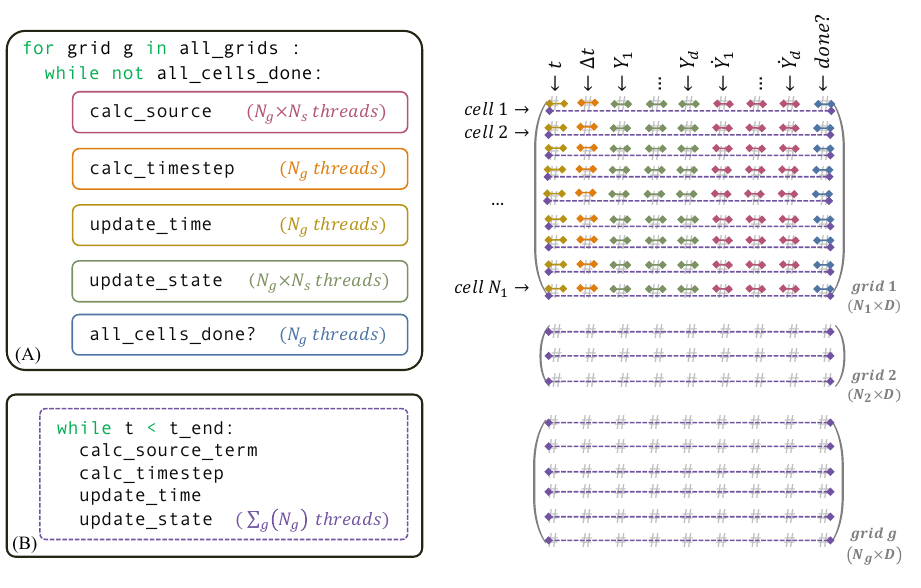}
    \caption{Schematic of the different parallelization strategies implemented in (A) the baseline kinetics code and (B) the optimized kinetics code. Each boxed step in the pseudocodes represents a different GPU kernel launched with the specified number of threads. The different line colors and styles overlayed on the matrix of grid data represent the data elements that the GPU threads of a corresponding GPU kernel write to. One line equals one GPU thread. In pseudocode (B), a single GPU kernel is launched with a thread count equal to the cell count across all grids.}
    \label{fig_par_strategies}
\end{figure}

To reduce kernel launch costs in both bulk and sparse integration, we launch a single kernel across all cells from all grids. This design choice sacrifices some parallelization strategies over others. To guide this explanation, refer to Figure~\ref{fig_par_strategies}, which showcases two implementations of kinetics on GPUs. In a single-grid solver with contiguous memory layout, we could parallelize across both cells and species ($N_c \times N_s$ threads), enabling the use of GPU-optimized libraries like cuBLAS \citep{NVIDIA_cuBLAS} for matrix operations, as seen in Figure~\ref{fig_par_strategies}. This approach was utilized by \cite{barwey2021neural} to achieve superior performance when working with (i) a single medium to large grid (see \ref{apx_naive_vs_cublas_dgemm}) or (ii) a low grid count in a multi-grid setting (see \ref{apx_grid_behavior}). This multi-kernel strategy also eliminates warp divergence by checking whether all cells are complete before continuing the \texttt{while} loop iteration.

However, the multi-kernel strategy becomes expensive in multi-grid settings with larger grid counts and more disparate grid sizes. This is caused by the kernel launch costs associated with launching multiple sequences of kernels for each grid. Moreover, cuBLAS performs poorly on small grids (see Figure~\ref{fig_naive_vs_cublas_dgemm}). Since our AMR applications typically process hundreds of small grids over hundreds of time steps (see \ref{apx_active_cell_count_behavior} and \ref{apx_grid_behavior}), the cumulative effects of kernel launch overhead and poor cuBLAS performance on small grids outweigh any benefits from cuBLAS optimizations or the elimination of warp divergence. 

This motivated the development of a single-kernel approach, where each thread is mapped to a single cell and performs serial iterations over its elements, as shown in Figure~\ref{fig_par_strategies}. Our solution sacrifices cuBLAS and multi-dimensional parallelization in favor of a single-kernel strategy that processes all grids simultaneously using a cell index map. While this reduces arithmetic efficiency per kernel (discussed in Section~\ref{sec_results_roofline}) and may introduce warp divergence as cells complete integration at different rates, the elimination of kernel launch overhead yielded substantial overall performance gains, as demonstrated in Section~\ref{sec_results_multi_gpus}.

\begin{algorithm}[H]
    \caption{Naive Kinetics Pseudo-Code for GPUs}
    \label{alg_naive_kinetics_solver}
    \begin{algorithmic}[1]
    \For{Fab \textbf{in} multiFab}
        \ParFor{cell $i$ \textbf{in} Fab}
            \State \textcolor{comment}{// INITIALIZATION}
            \State $t_i \leftarrow 0$ \textcolor{comment}{// initialize time}
            \State $k_i \leftarrow 0$ \textcolor{comment}{// initialize step counter}
            \If{$T_i$ $<$ $T_{\text{reaction\_min}}$ \textbf{or} \textcolor{function}{is\_solid}(cell $i$)}
                \State $t_i \leftarrow t_\text{final}$ \textcolor{comment}{// cell will not be integrated}
            \EndIf
            \State
            \State \textcolor{comment}{// MAIN INTEGRATION LOOP}
            \While{$t_i < t_{\text{final}}$ \textbf{and} $k_i < K_{\text{max}}$}
                \State $\vec{S}_i \leftarrow$ \textcolor{function}{calc\_source\_terms}() \textcolor{comment}{// $N_s$-dimensional source term vector}
                \State $\Delta t_i \leftarrow$ \textcolor{function}{calc\_timestep\_size}()
                \State $t_i \leftarrow t_i + \Delta t_i$ \textcolor{comment}{// update time, state, and step counter}
                \State $\vec{Y}_i \leftarrow \vec{Y}_i + \Delta t_i \times \vec{S}_i$
                \State $k_i \leftarrow k_i + 1$
            \EndWhile
        \EndParFor    
    \EndFor
    \end{algorithmic}
\end{algorithm}

\begin{algorithm}
    \caption{Optimized Integration with Bulk and Sparse Processing}
    \label{alg_optimized_integration}
    \begin{algorithmic}[1]
    \State \textcolor{comment}{// SECTION 1: COUNT CELLS REQUIRING REACTION}
    \State counter $\leftarrow$ \textcolor{number}{0}
    \ParFor{cell $i$ \textbf{in} multiFab}
        \State $t_i \leftarrow 0$ \textcolor{comment}{// initialize time}
        \If{$T_i$ $<$ $T_{\text{reaction\_min}}$ \textbf{or} \textcolor{function}{is\_solid}(cell $i$)}
            \State $t_i \leftarrow t_\text{final}$ \textcolor{comment}{// cell will not be integrated}
        \Else
            \State counter $\leftarrow$ \textcolor{function}{atomic\_plus\_one}(counter) \textcolor{comment}{// add to active cell count}
        \EndIf
    \EndParFor
    \State $N_{\text{active}} \leftarrow$ \textcolor{function}{copy\_device2host}(counter) \textcolor{comment}{// memory copy to host}

    \State
    \State \textcolor{comment}{// SECTION 2: BULK INTEGRATION ON ALL CELLS}
    \While{$N_{\text{active}} > N_{\text{active}}^*$}
        \State counter $\leftarrow$ \textcolor{number}{0}
        \ParFor{cell $i$ \textbf{in} multiFab}
            \State $k_i \leftarrow 0$ \textcolor{comment}{// reset step counter}
            \State \textcolor{function}{timeIntegration}(cell $i$, $t_{\text{final}}$, $K_{\text{max}}$)
            
            \If{$t_i < t_{\text{final}}$}
                \State counter $\leftarrow$ \textcolor{function}{atomic\_plus\_one}(counter) \textcolor{comment}{// add to active cell count}
            \EndIf
        \EndParFor
        \State $N_{\text{active}} \leftarrow$ \textcolor{function}{copy\_device2host}(counter) \textcolor{comment}{// memory copy to host}
    \EndWhile
    
    \State
    \State \textcolor{comment}{// SECTION 3: SPARSE INTEGRATION ON REMAINING ACTIVE CELLS}
    \If{$N_{\text{active}} > \textcolor{number}{0}$}
        \State active\_ids $\leftarrow$ \textcolor{function}{allocate\_vector}(size=$N_{\text{active}}$) \textcolor{comment}{// to store locations of active cells}
        \State id $\leftarrow$ \textcolor{number}{0}
        \ParFor{cell $i$ \textbf{in} multiFab}
            \If{$t_i < t_{\text{final}}$}
                \State id $\leftarrow$ \textcolor{function}{atomic\_plus\_one}(id)
                \State active\_ids[id] $\leftarrow$ $i$ \textcolor{comment}{// store location of active cell}
            \EndIf
        \EndParFor
        
        \ParFor{$i$ \textbf{in} active\_ids}
            \State $k_i \leftarrow 0$ \textcolor{comment}{// reset step counter}
            \State \textcolor{function}{timeIntegration}(multiFab[$i_\text{active}$], $t_{\text{final}}$, $K_{\text{max}}$=\textcolor{number}{$10^5$})
        \EndParFor
    \EndIf
    \end{algorithmic}
\end{algorithm}

\subsection{Convection Routine Optimizations}

A key optimization involved implementing dynamic tiling strategies for the convection routines. In AMReX, GPU execution is orchestrated through a hierarchical decomposition in which each refinement level is divided into boxes (or grids) assigned to thread blocks for parallel execution. By default, when tiling is not explicitly enabled, AMReX launches one GPU thread block per box. This strategy minimizes kernel launch overhead and maximizes the amount of data handled per block, but it could result in suboptimal hardware utilization when the box dimensions are large, anisotropic, or uneven across ranks. Large grids increase register pressure and shared-memory usage within a single thread block, reducing occupancy and the number of active warps per streaming multiprocessor. 

The problem becomes more pronounced when the numerical method requires several temporary variables defined at the grid level. Such temporary variables which are allocated per box, consume device memory in direct proportion to the grid volume. Consequently, large boxes not only inflate the overall memory footprint but also increase register usage when these variables are accessed within GPU kernels. The resulting register pressure can lead to spilling into local memory, which further diminishes occupancy and degrades effective memory throughput.

To address these inefficiencies, we implemented manual tiling, in which each grid is subdivided into smaller computational blocks whose dimensions are determined dynamically from the average box size within the AMR hierarchy. This strategy improves GPU occupancy and cache reuse, enhances latency hiding, and increases overall device utilization. Although this approach does not ensure that all temporary or shared variables remain within on-chip memory limits, the smaller per-tile working set generally reduces register pressure and mitigates local-memory spills relative to the default per-box execution. 

A further advantage of this tiling strategy is its positive impact on load balancing across CUDA streams. When tiles have comparable dimensions, the workload distribution becomes more uniform, reducing execution imbalance among concurrent streams or across streaming multiprocessors. This uniformity enables the GPU scheduler to overlap compute and memory operations more effectively, sustaining high occupancy even when computational cost varies spatially—for instance, in regions with embedded boundaries. This strategy is particularly effective for convection routines, where the underlying kernels are strongly memory-bound (see Figure~\ref{fig_rooflines_all}). Compared with reducing the global grid size which increases the communication overhead, the dynamic tiling approach provides finer control over workload distribution without incurring the additional cost associated with managing extra ghost cells.

\section{Setup for Performance Analysis}\label{sec_setup_for_performance_analysis}

\subsection{Setup of the Hardware, Software, and Performance Metrics}

We obtain all results reported in Section~\ref{sec_results} by running the solver on a cluster with NVIDIA H100 GPUs \citep{h100specs}. The GPUs are peer-to-peer connected using 18 4th-generation NVLINKs, with a total peak bandwidth of 900 GB/s, supporting simultaneous bidirectional communication. Empirical bandwidth tests using sample CUDA scripts \citep{nvidia_bandwidthTest} measured net unidirectional data transfer rates of  $\approx 478$ GB/s between any two GPUs. GPUs are connected to Intel(R) Xeon(R) Platinum 8468 CPUs using PCIe 5th Gen communication technologies, with peak bandwidths of 64 GB/s. The simulations were executed with each CPU process bound to a single GPU, as required by the AMReX GPU implementation. Therefore, the number of CPUs used for a simulation equals the number of GPUs.

We compiled two versions of the executable for a given set of inputs. The first version is compiled with level 3 optimizations using the \texttt{GNU} suite of compilers. The second version is a suboptimal compilation with debugging enabled, allowing for the collection of wall timing for various code regions. We also used the debugging version to collect detailed GPU kernel metrics using NVIDIA Nsight Compute \citep{NVIDIANsightCompute} to derive quantities for the roofline model. These metrics are detailed in \cite{Yang2020}.

\subsection{Code Versions}\label{sec_code_versions}

To evaluate the impact of our optimization strategies, we compared two implementations of the solver throughout the analysis in Section~\ref{sec_results}. These are summarized in Table~\ref{tab_code_comparison}. The baseline implementation is the initial GPU port of our AMReX-based combustion solver, which utilizes cuBLAS for matrix operations within kinetics, but does not include the memory, chemistry, and convection optimizations described in Section~\ref{sec_gpu_implementation}. The optimized version incorporates these optimizations. Performance comparisons between these versions demonstrate the effectiveness of our optimization approach across multiple GPUs (Section~\ref{sec_results_multi_gpus}), runtime breakdowns (Figures~\ref{fig_runtime_pies}~and~\ref{fig_runtime_breakdown}), and roofline analysis (Section~\ref{sec_results_roofline}).

\begin{table}[htbp]
    \centering
    \caption{Comparison of key features between baseline and optimized 
             GPU-based combustion solver versions.}
    \label{tab_code_comparison}
    \renewcommand{\arraystretch}{1.4}
    \begin{tabular}{|p{0.25\textwidth}|p{0.33\textwidth}|p{0.33\textwidth}|}
        \hline
        \textbf{Feature} & \textbf{Baseline Version} & \textbf{Optimized Version} \\
        \hline
        \textbf{Memory Management} & 
        Default AMReX memory allocation; Managed memory enabled & 
        Unchanged \\
        \hline
        \textbf{Data Storage} & 
        Default AMReX column-major with minimal optimization & 
        Default AMReX column-major with improved memory 
        coalescence patterns \\
        \hline
        \textbf{Kernel Launch in Chemistry} & 
        Multiple cuBLAS-enabled kernels per grid per time step; Excessive launch overhead & 
        One kernel across all grids every few time steps; cuBLAS removed; Reduced launch overhead \\
        \hline
        \textbf{Active Cell Handling in Chemistry} & 
        All cells integrated regardless of completion status & 
        Bulk integration followed by sparse integration on the remaining 
        active cells \\
        \hline
        \textbf{Convection} & 
        Standard grid processing without tiling & 
        Dynamic tiling strategies; Adaptive tile sizing based on average grid size \\
        \hline
    \end{tabular}
\end{table}

\subsection{Description of The Combustion Applications}\label{sec_combustion_applications}

Here, we present the performance of the solver across three distinct combustion applications with increasing degrees of freedom, two 2D detonation cases, and one 3D jet in a supersonic crossflow. The cases are representative of the computational challenges commonly encountered in high-speed reacting flows. They collectively stress different aspects of combustion solvers -- shock capturing, stiff chemistry, AMR efficiency, multiscale physics, and more. Here we provide a brief description of each case.

\begin{figure}[h]
    \centering
    \includegraphics[width=\textwidth]{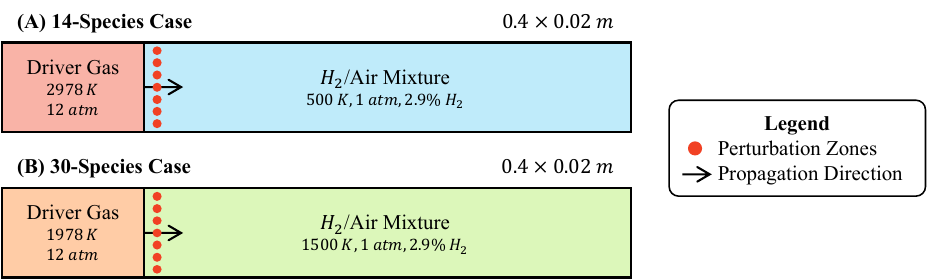}
    \caption{Schematic of the initial and boundary conditions for the 2D detonation cases. (A) Cold detonation in a hydrogen-air mixture using a reduced 14-species mechanism by \cite{mevel2009hydrogen}. (B) Hot detonation in a hydrogen-air mixture using a 30-species mechanism by \cite{smith2020foundational}. All walls are non-slip and adiabatic. The driver is a high-temperature, high-pressure gas that propagates into a stagnant hydrogen-air mixture. Perturbation zones reside slightly downstream of the driver to accelerate ignition and facilitate the formation of cellular detonation structures.}
    \label{fig_detonation_cases}
\end{figure}

\textbf{2D Detonation Cases:} We studied two variants of a planar detonation wave propagating into a hydrogen-air mixture. The first variant employs the 14-species mechanism by \cite{mevel2009hydrogen}, while the second uses the reduced 30-species mechanism by \cite{smith2020foundational}. These cases are based on the work by \cite{yungster2005structure} and are simple enough for performance analysis yet still feature stiff chemistry encountered in many combustion simulations. Moreover, the presence of discontinuities and thin reaction zones demands resolving multiple length and time scales, which is well-suited for the adaptive mesh refinement. As schematized in figure~\ref{fig_detonation_cases}, each case consists of a 2D domain with a high-temperature, high-pressure gas on the left propagating into a stagnant hydrogen-air mixture. Perturbation zones reside slightly downstream of the driver to accelerate ignition and facilitate the formation of cellular detonation structures.

\textbf{Jet in Supersonic Crossflow (JISCF):} This application simulates a sonic Ethylene jet injected into a high-temperature Mach 2.48 crossflow, representing conditions relevant to high-speed propulsion systems. The interested reader may find details of the case in \cite{SHARMA2024105295}. The configuration features complex physics, including compressible turbulence, shock-turbulence interactions, fuel-air mixing, and finite-rate chemistry effects. This case employs a reduced, 30-species, 231-reaction mechanism proposed by \cite{smith2020foundational} and presents significant computational challenges due to the three-dimensional nature of the flow, the presence of multiple shocks, and the strong coupling between turbulent mixing and chemical kinetics in the high-speed regime.

\subsection{Choice of Parameters}\label{sec_choice_of_parameters}

The parameters we selected for performance assessment on multiple GPUs are reported in Table~\ref{tab_case_params}. We selected these values for performance analysis due to their tractability and relevance to practical cases. For instance, we used 2 AMR levels for all three combustion applications to highlight the AMR features of the solver; however, the number of AMR levels required depends on the application. Unlike the JISCF case, the 2D detonation cases are small enough to run for several time steps without extensive wall times. The selected chemical mechanisms by \cite{mevel2009hydrogen} (14-species) and \cite{smith2020foundational} (30-species) are chosen to trace the variation in performance with increasing chemical complexity in well-established chemistry models.

\begin{table*}[htbp]
    \centering
    \caption{Parameters by case for performance assessment. Results shown in Figures~\ref{fig_runtime_breakdown}~and~\ref{fig_runtime_pies}.}
    \label{tab_case_params}
    \renewcommand{\arraystretch}{2}  
    \begin{tabular*}{\textwidth}{@{\extracolsep{\fill}}|r|c|c|c|}
        \hline
        \hline
        \textbf{} & \textbf{2D Detonation} & \textbf{2D Detonation} & \textbf{JISCF\textsuperscript{*}} \\
        \hline
        \hline
        \textbf{Mechanism} & \cite{mevel2009hydrogen} & \multicolumn{2}{c|}{\cite{smith2020foundational}} \\
        \hline
        \textbf{Coarse Cells} & \multicolumn{2}{c|}{ 40960 $\times$ 2048 } & 768 $\times$ 128 $\times$ 128 \\
        \hline
        \textbf{GPUs} & \multicolumn{3}{c|}{16} \\
        \hline
        \textbf{Time Steps} & \multicolumn{2}{c|}{250} & 30 \\
        \hline
        \textbf{AMR Levels} & \multicolumn{3}{c|}{2} \\
        \hline
        \textbf{Code Version} & \multicolumn{3}{c|}{Baseline, Optimized} \\
        \hline
        \hline
        \multicolumn{4}{l}{\textsuperscript{*}More details of this case can be found in \cite{SHARMA2024105295}} \\
    \end{tabular*}
\end{table*}

\section{Performance Results and Discussion}\label{sec_results}

\subsection{Multi-GPU Performance}\label{sec_results_multi_gpus}

\begin{figure}[H]
    \centering
    \includegraphics[width=0.9\textwidth]{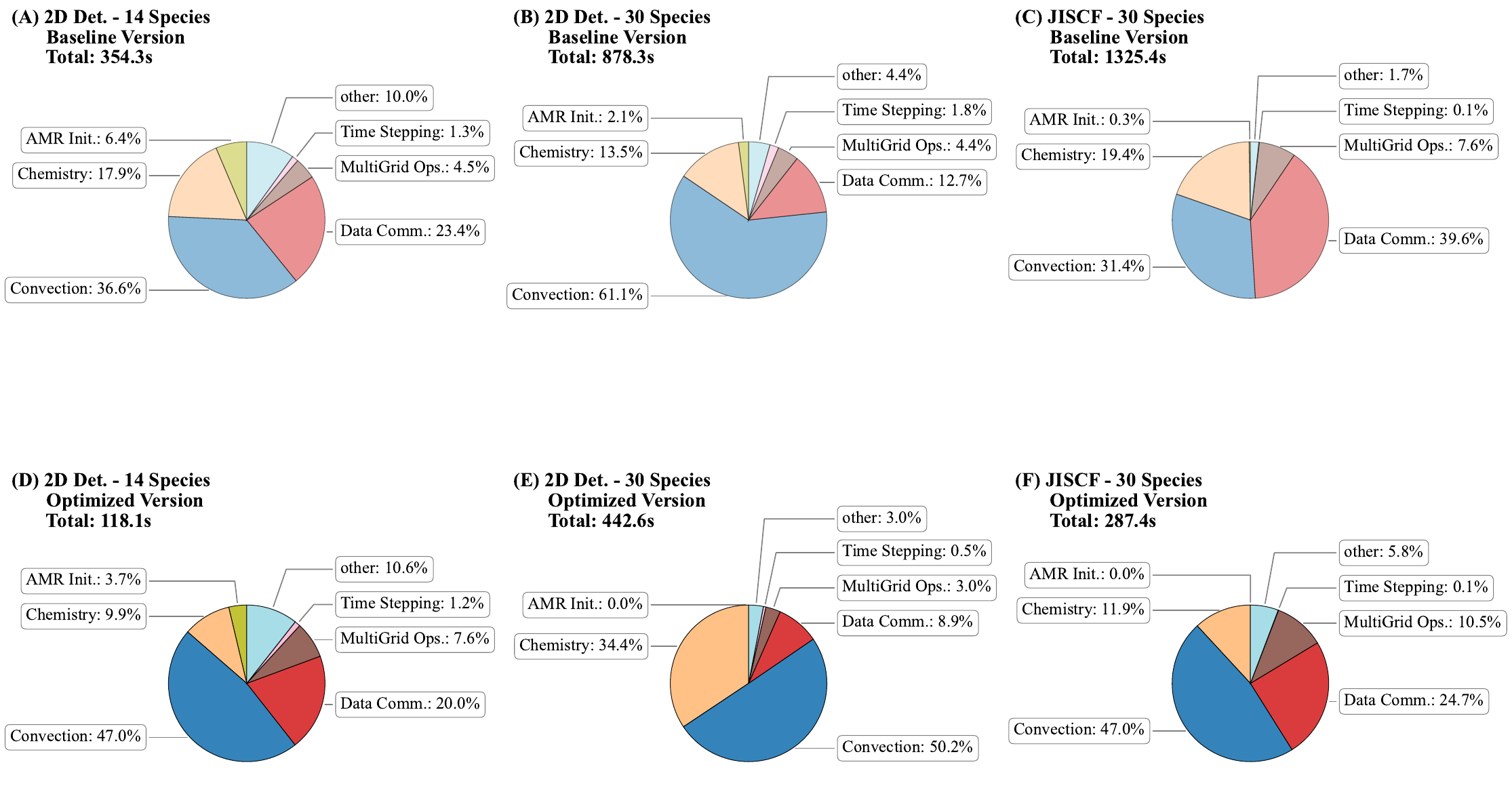}
    \caption{Relative runtime breakdown across different code regions at 16 GPUs for three combustion applications: 2D detonation with 14 species mechanism (panels A/D), 2D detonation with 30 species mechanism (panels B/E), and JISCF (panels C/F). Panels A, B, and C show the baseline version results, while panels D, E, and F show the optimized version results.}
    \label{fig_runtime_pies}
\end{figure}

\begin{figure}[H]
    \centering
    \includegraphics[width=0.9\textwidth]{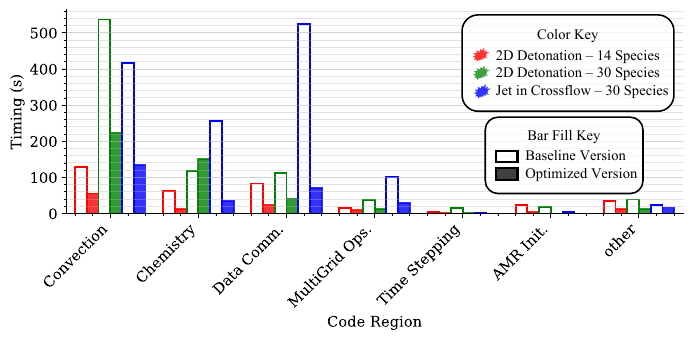}
    \caption{Absolute runtime breakdown across different code regions at 16 GPUs for three combustion applications: 2D detonation with 14 species mechanism (red bars), 2D detonation with 30 species mechanism (green bars), and JISCF (blue bars).}
    \label{fig_runtime_breakdown}
\end{figure}

This section examines the multi-GPU performance of the solver across the three combustion applications. Figure~\ref{fig_runtime_pies} shows the total runtime for each case using the baseline and optimized code versions, along with the fractional breakdown of the runtime. The 14-species 2D det., 30-species 2D det., and JISCF cases achieved a $\sim 3 \times$, $\sim 2 \times$, and $4.6 \times$ speedup, respectively.

The superior performance gain for JISCF is evident in the absolute timing breakdown in Figure~\ref{fig_runtime_breakdown}, where a significant improvement is observed across the convection, chemistry, and communication routines. This behavior is expected, given the computational demands of the JISCF case and the opportunities for performance improvement. For instance, stiff chemistry may lead to many cells completing the reaction in a couple of time steps, with very few cells remaining active for tens or hundreds of time steps, as seen in Figure~\ref{fig_reaction_clusters}. Consequently, we observe significant performance gains due to the efficient integration of only the active cells. For the convection routines, the optimized version benefits significantly from dynamic tiling strategies that improves load balancing and enables better utilization of GPU resources. Communication improvements stem from reduced synchronization overhead between GPU kernels and more efficient data movement patterns enabled by the optimized memory management strategies.

The 14-species 2D detonation also showcases a significant overall performance improvement and a decrease in runtime across all code regions. The speedup is attributed to the improved load balancing in the convection routines and reduced communication overhead. We also observe a significant decrease in runtime for chemistry from the combined effect of (1) efficient integration on only active cells, similar to the JISCF case, and (2) kernel fusion across multiple grids. Regarding the performance effect of kernel fusion, we see a performance gain because the minimum number of grids needed to achieve performance gains from kernel fusion -- in Figure~\ref{fig_runtime_vs_n_grids}, 15 grids for a 14-species mechanism assuming 1 million cells across all grids running on 1 GPU -- is reached across all AMR levels (see Figure~\ref{fig_grid_size_distribution}).

The 30-species 2D detonation case exhibits a modest yet appreciable improvement. While speedup is still attributed to the reasons discussed previously, the larger mechanism is more computationally expensive, as each GPU thread needs to iterate over more species and reactions to compute reaction rates, source terms, and updated species states. In Figure~\ref{fig_runtime_breakdown}, the optimized version of the chemistry subroutine actually performs worse than the baseline version, with a $\sim 30\%$ increase in runtime. This performance drop is likely because the minimum number of grids needed to achieve performance gains from kernel fusion -- in Figure~\ref{fig_runtime_vs_n_grids}, 150 grids for a 30-species mechanism assuming 1 million cells across all grids running on 1 GPU -- is not achieved across any AMR level (see Figure~\ref{fig_grid_size_distribution}). While a similar process is at play for the 30-species JISCF, we still see a dramatic improvement in the chemistry routine due to the effect of efficient integration on only the active cells. What this reveals is a complex interplay between the number of grids and the drop-off trend in active cells per grid, that are problem-dependent. In future work, we will explore the use of prediction models to guide the selection of simulation parameters, including different algorithms for various subroutines, to optimize the performance of the chemistry routines and the solver as a whole throughout the simulation.

\subsection{Roofline Performance}\label{sec_results_roofline}

\begin{figure}[h]
    \centering
    \includegraphics[width=\textwidth]{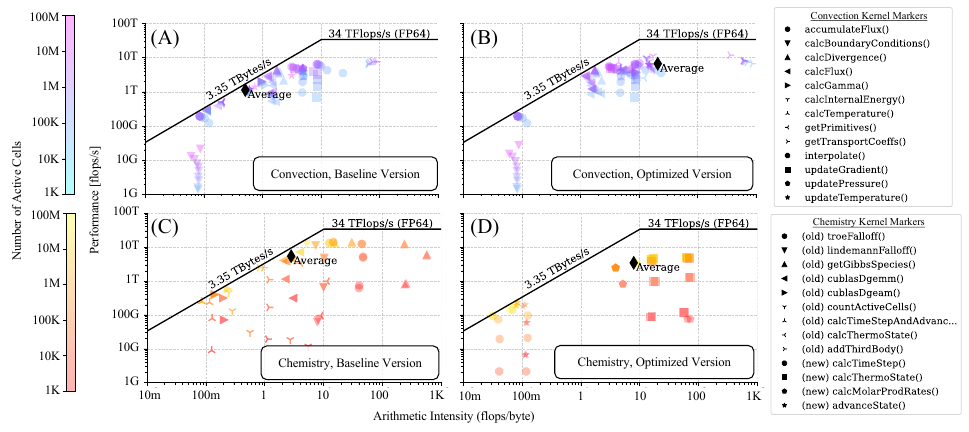}
    \caption{Roofline performance with varying active cell counts of GPU kernels across the convection (panels A and B) and chemistry (panels C and D) routines. Kernels in the baseline (panels A and C) and optimized (panels B and D) versions of the code are shown.}
    \label{fig_rooflines_all}
\end{figure}

Figure~\ref{fig_rooflines_all} presents roofline performance results for both convection and chemistry routines across both versions of the solver with varying active cell counts. More markers tending towards the top-right of the roofline indicate better performance. A rightward shift indicates a higher arithmetic intensity (AI), meaning the kernel performs more floating-point operations per byte of data accessed. An upward shift indicates faster calculations per second. Each subplot also shows the average kernel marker as a black diamond. We calculated its AI by dividing the sum of FLOPs across all kernels by the sum of bytes accessed by the kernels. Similarly, its performance is the sum of FLOPs across all kernels divided by the sum of run times. We use the average kernel marker to measure the overall roofline performance of the kernels that are displayed in a given subplot.

The average kernel marker in the convection routine shifts right and upwards from Panel A to Panel B, demonstrating a dramatic improvement in both AI and performance by $\sim 10 \times$. This improvement is partially driven by the \texttt{getTransportCoeffs()} and \texttt{calcFlux} kernels, among others. We rewrote these kernels to improve register usage, increase reuse of pre-calculated values, and enhance memory access patterns. Notably, the \texttt{calcBoundaryConditions} kernel remains poor-performing and memory-bound in both versions of the code. We expect this behavior, given that this kernel primarily performs assignment operations and lacks floating-point operations. Tiling optimizations also contribute significantly to the $\sim 10\times$ improvement in AI observed. The approach scales well across multiple GPUs, as each GPU processes appropriately sized tiles that maintain high occupancy regardless of the underlying grid distribution, resulting in the performance improvements seen in the convection routines in Figure~\ref{fig_runtime_breakdown}.

The average AI in the chemistry routine improves by $\sim 4 \times$ while the performance drops by $\sim 40 \%$, which is an expected outcome. One key dilemma encountered in implementing Algorithm~\ref{alg_optimized_integration} is the tradeoff between (A) multiple kernels, each performing a subset of steps for the source term computation and state advancement, and (B) a single kernel that performs all steps. Approach (A) is highly effective in single-grid settings where all cell data is stored contiguously in memory. As detailed in Section~\ref{sec_cublas_par_dims}, each kernel's threads can parallelize over as many dimensions of the data as possible, and some kernels may leverage cuBLAS, a GPU-optimized version of the Basic Linear Algebra Subroutines (BLAS) library \citep{blackford2002updated}. However, this approach is not suitable in multigrid settings where iter-grid data may not reside contiguously. Each grid would require its own sequence of kernel launches, leading to excessive kernel launch costs with large grid counts. Moreover, sparse integration would require copying active cell data to and from a contiguous chunk of memory to support cuBLAS and multi-dimensional parallelization. These memory copies become costly with large grid counts. In approach (B), we lost cuBLAS optimizations and multi-dimensional data parallelization; parallelization is now limited to the cell-wise dimension. Figure~\ref{fig_rooflines_all} shows this penalty. However, we gained the ability to launch a single kernel across all grids by using a cell index map, which is key to the performance improvements observed in Figures~\ref{fig_runtime_vs_n_grids},~\ref{fig_runtime_pies},~and~\ref{fig_runtime_breakdown}.

The roofline analysis in Figure~\ref{fig_rooflines_all} confirms the expected performance characteristics of our kernel fusion strategy (see last paragraph of Section~\ref{sec_chemistry_algorithm_optimization}). While the average AI in the chemistry routine improves by $\sim 4 \times$, the performance drops by $\sim 40\%$ per kernel, reflecting the computational tradeoffs inherent in our single-kernel approach. This performance penalty at the kernel level is justified by the elimination of kernel launch overhead when processing the hundreds of grids typical in our AMR applications (see \ref{apx_grid_behavior}). The net result is the substantial overall speedups observed in Figures~\ref{fig_runtime_pies}~and~\ref{fig_runtime_breakdown}, demonstrating that system-level optimization can outweigh individual kernel inefficiencies in multigrid combustion simulations.

The color coding in Figure~\ref{fig_rooflines_all} indicates better performance at higher active cell counts, consistent with the notion that applications with a large volume of highly-parallel work are well-suited for GPU acceleration. This trend explains why we observe performance improvements in Figure~\ref{fig_runtime_vs_n_grids} when launching a single large kernel across multiple grids, compared to launching many smaller kernels, one per grid. What this also reveals is that performance gains or losses in a roofline model are not always indicative of the performance of the entire application. Examining the overall roofline performance of the new convection and chemistry routines, we still see clusters of kernels that are memory-bound (underneath the slanted solid black line in Figure~\ref{fig_rooflines_all}). Only continued rewritings and optimizations may yield further performance improvements. In particular, utilizing shared and constant memory, exploring various memory layouts, fusing multiple kernels, and redesigning algorithms may be future avenues to explore.

\section{Conclusions}\label{sec_conclusion}

We have presented a high-performance GPU-accelerated compressible reacting flow solver built on the AMReX framework, optimized explicitly for H100 GPUs. The solver addresses three critical challenges for GPU performance in multiscale combustion simulations: memory access patterns, load balancing across multiple GPUs, and computational workload variability arising from highly localized chemical reactions.

\subsection{Key Contributions}

The primary contributions of this work include: (1) a GPU-optimized implementation of patch-based adaptive mesh refinement for compressible reacting flows that maintains the algorithmic advantages of AMReX while achieving significant performance improvements on modern GPU architectures; (2) an efficient bulk-sparse integration strategy for chemical kinetics that minimizes thread divergence and kernel launch overhead while exploiting the multiscale nature of combustion chemistry \citep{barwey2021neural}; (3) optimized memory management strategies; and (4) optimized load distribution via tiling in the convection routine.

The solver demonstrates the feasibility of achieving high-performance combustion simulations on GPU clusters while maintaining the adaptive capabilities essential for resolving multiscale reacting flows. By addressing GPU-specific performance bottlenecks within the established AMReX framework, this work provides a pathway for existing combustion codes to leverage modern exascale computing architectures \citep{reed2015exascale}.

\subsection{Solver Performance}

The optimized solver demonstrates substantial performance improvements across three representative combustion applications. For the $14$-species hydrogen-air detonation case, we achieved approximately $3\times$ speedup over the initial GPU implementation, while the 30-species detonation case showed $2\times$ improvement. We observed the most significant gains for the jet in supersonic crossflow (JISCF) application, where the solver achieved a $5\times$ speedup, attributed to the efficient handling of highly variable chemical activity characteristic of turbulent reacting flows.

Weak scaling analysis across $1-96$ H100 GPUs demonstrates near-ideal scaling behavior for all test cases, with performance remaining essentially constant as both problem size and GPU count increase proportionally. The roofline analysis reveals substantial improvements in arithmetic intensity for convection routines ($\sim 10\times$ increase) and chemistry routines ($\sim 4\times$ increase), indicating more efficient utilization of GPU memory bandwidth. 

The chemistry optimization strategy proved effective for cases with high stiffness, where the bulk-sparse integration approach dramatically reduces computational overhead by avoiding unnecessary work on inactive cells. This strategy mirrors the benefits observed in previous GPU chemistry implementations \citep{barwey2021neural}, but extends these concepts to the multigrid AMR context, which is essential for practical combustion simulations.

\subsection{Limitations and Future Directions}

While the current implementation achieves significant performance improvements, several opportunities remain for further optimization. For instance, the chemistry kernels, despite improved arithmetic intensity, still exhibit memory-bound behavior in many cases, suggesting potential benefits from shared memory utilization, alternative memory layouts, and additional kernel fusion strategies. The load-balancing approach, while effective for the tested configurations, may require adaptation for extreme variations in chemical activity or highly non-uniform grid distributions.

Future work will focus on (1) integration of machine learning techniques for predictive performance optimization and automatic code generation, building on recent advances in GPU performance modeling \citep{Chennupati2021-dw, Arafa2021-sf}; (2) exploration of CUDA dynamic parallelism and cooperative thread arrays to better handle workload imbalance at the thread level; (3) investigation of unified memory management strategies for next-generation GPU architectures that may reduce the complexity of manual memory management; and (4) development of architecture-specific optimization strategies that can automatically adapt solver parameters based on hardware characteristics and application requirements.

The continued evolution of exascale systems and programming models presents ongoing opportunities to improve performance further. As exascale systems increasingly rely on GPU acceleration, the optimization strategies demonstrated in this work provide a foundation for developing next-generation combustion simulation capabilities that can fully exploit these advanced computing platforms.

\appendix

\section{Verification of the Kinetics Solver}

To verify the accuracy of our GPU-accelerated kinetics implementation, we compared source terms and ignition delay times with those from the established CPU-based Cantera library \citep{cantera} using three chemical mechanisms. Note that, in addition to verifying the accuracy of the mechanisms used in this study, we also verify the GRI mechanism to confirm further the reliability of our optimization strategies for production combustion simulations.

\begin{figure}[H]
    \centering
    \includegraphics[width=0.85\linewidth]{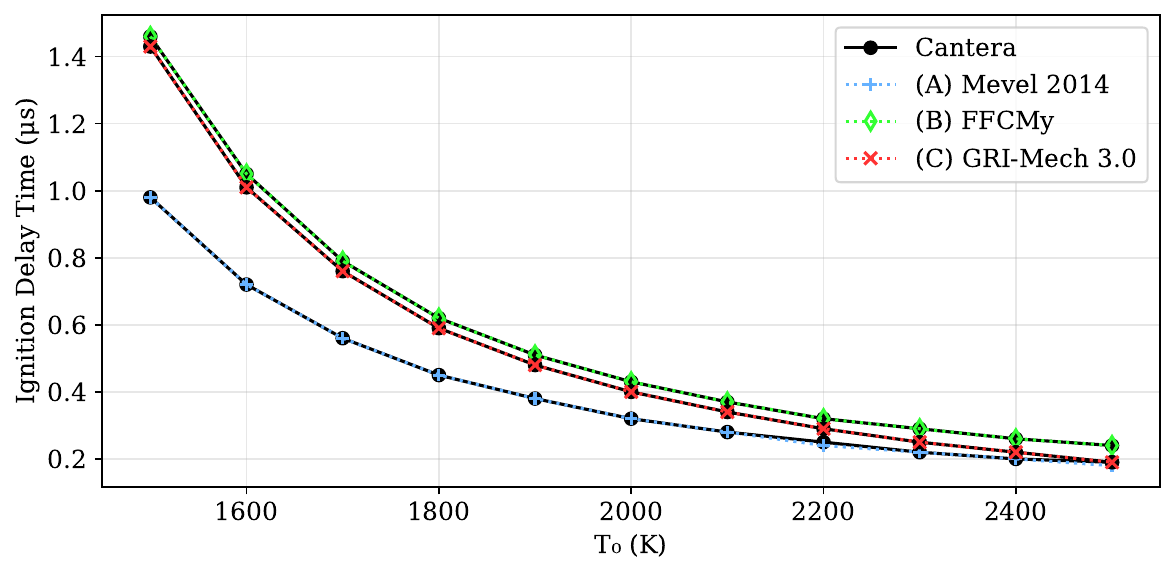}
    \includegraphics[width=\linewidth]{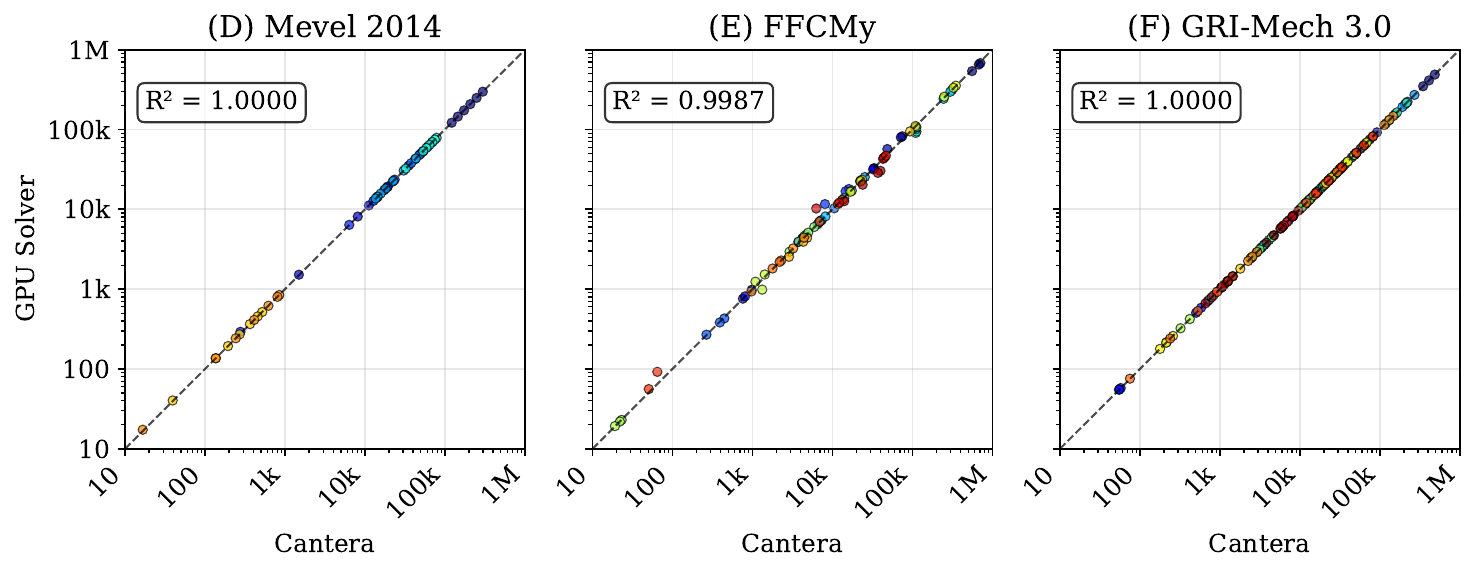}
    \caption{Accuracy verification of the kinetics solver. (A,B,C) Comparison of ignition delay times for hydrogen-air mixtures at stoichiometric conditions and 1 atm pressure between Cantera reference solutions (solid circles) and our GPU solver using three chemical mechanisms: (A) the 14-species Mevel mechanism \citep{mevel2009hydrogen}, (B) the reduced 30-species FFCMy mechanism \citep{smith2020foundational}, and (C) the 53-species GRI-Mech 3.0 \citep{smithgri} mechanism. (D,E,F) Species molar production rate correlations between GPU solver calculations (y axis) and Cantera calculations (x axis) for the same chemical mechanisms. The dashed line represents perfect agreement. Different colors represent different chemical species. $R^2$ values are also reported.}
    \label{fig_verification}
\end{figure}

Figure~\ref{fig_verification} shows ignition delay time comparisons for hydrogen-air mixtures starting at stoichiometric conditions and 1 atm pressure across a range of initial temperatures from 1500K to 2500K for three mechanisms. The solver accurately captures the characteristic Arrhenius temperature dependence across all mechanisms with near-perfect agreement against Cantera reference solutions. 

Figure~\ref{fig_verification} presents scatter plots comparing species production rates between our GPU solver and Cantera. Species production rates were computed from randomly sampled temperatures and mass fractions at standard pressure. The correlations demonstrate exceptional accuracy, with unity $R^2$ values for the Mevel and GRI mechanisms and $R^2 = 0.9987$ for the FFCMy mechanism. We observe minor discrepancies for the FFCMy mechanism; these are attributed to compiler optimization differences and numerical precision effects for minor species, which do not affect the primary reaction pathways controlling ignition behavior, as evidenced by the perfect agreement in ignition delay predictions shown in Figure~\ref{fig_verification}.

\section{Active Cell Count Behavior in The Combustion Applications}\label{apx_active_cell_count_behavior}

The performance benefits of Algorithm~\ref{alg_optimized_integration} depend critically on the temporal evolution of active cell distributions across grids. To quantify this behavior and validate our optimization strategies, we instrumented the solver to track active cell counts throughout chemical integration and analyzed the resulting patterns using clustering techniques.

\subsection{Data Collection Methodology}

Active cell count data was collected by instrumenting the chemistry routine to output the number of active cells after each integration step for every grid within the \texttt{multiFAB} \texttt{MFIter} loop at each AMR level. Specifically, the code printed to the terminal: \texttt{Level <level>, FAB <fab\_ID>, t = <time>, step = <iteration>, n\_cells = <total\_cells>, n\_active = <active\_cells>}. This instrumentation captured the complete temporal evolution of chemical activity across all grids, levels, and time steps for each combustion application.

From this raw output, we constructed active cell count vectors for analysis. Let $\mathbf{a}_1$ $\mathbf{a}_2$, $\ldots$, $\mathbf{a}_g$, $\ldots$, $\mathbf{a}_G$ represent the active cell count vectors, where $G$ is the total number of grids across all levels and time steps. Each vector element $\mathbf{a}_g[i]$ corresponds to the number of active cells at chemistry iteration $i$ for grid $g$. To enable comparative analysis, all vectors were reshaped to contain $i_{\text{max}}$ elements, where $i_{\text{max}}$ represents the maximum number of integration steps taken across any grid. Grids requiring fewer than $i_{\text{max}}$ steps had their vectors padded with zeros to maintain consistent dimensionality.

The resulting dataset was analyzed using $k$-means clustering with the Wasserstein distance or Earth Mover's distance \citep{vaserstein1969markov} as the distance metric. We treated each active cell count vector as a point in $i_{\text{max}}$-dimensional space and partitioned the data into 15 non-overlapping clusters. The Wasserstein distance was selected after experimenting with various clustering algorithms and distance metrics, as it provided the most sensible groupings for visualization purposes by accounting for the temporal structure of the active cell decay patterns. We perform this procedure independently for each combustion application to capture case-specific chemical stiffness characteristics.

\subsection{Analysis of Active Cell Count Patterns}

\begin{figure}[H]
    \centering
    \includegraphics[width=\textwidth]{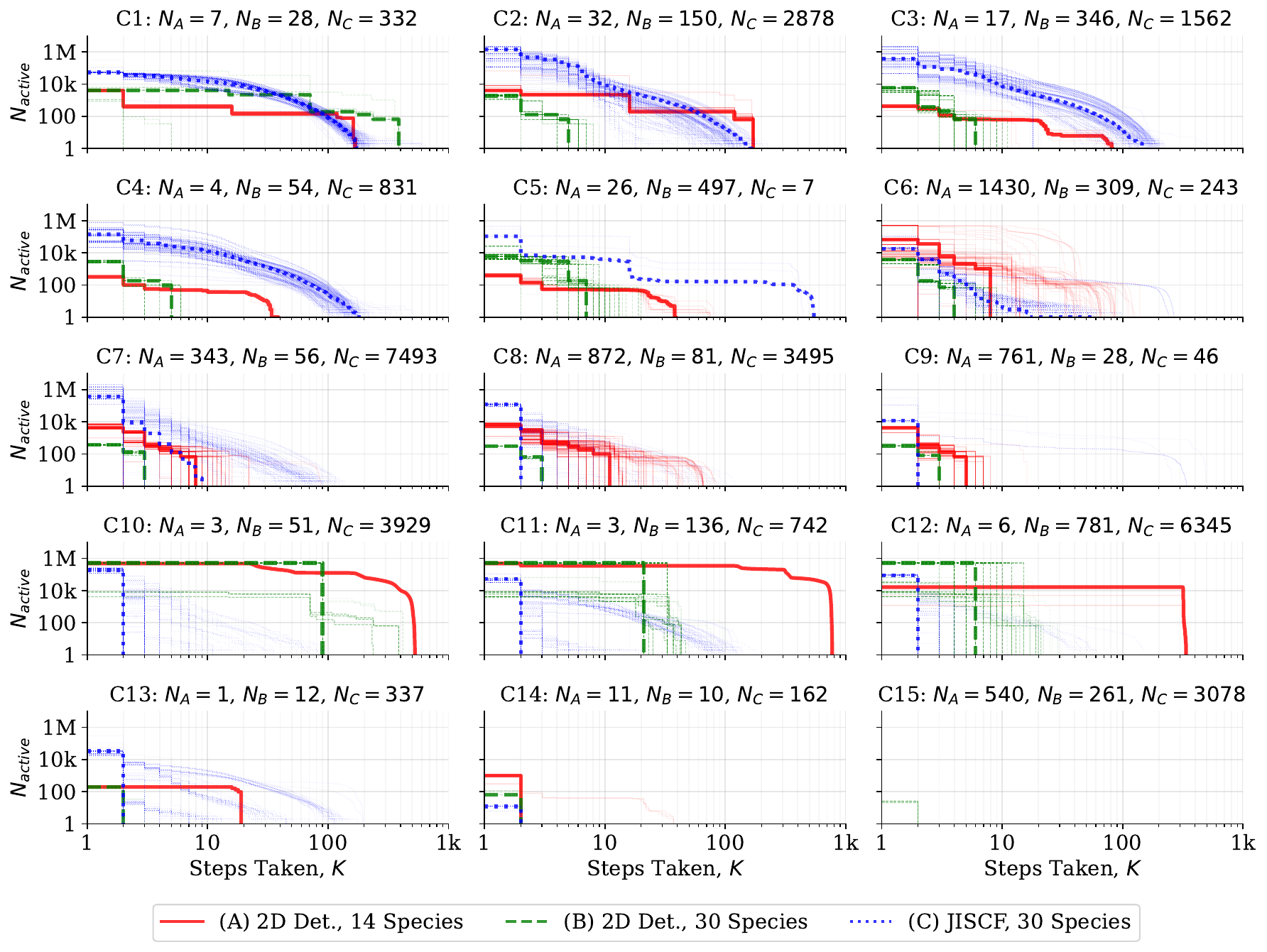}
    \caption{number of active cells remaining per grid after each time integration step for (A) 2D detonation with 14 species mechanism (blue, solid line), (B) 2D detonation with 30 species mechanism (red, dashed line), and (C) JISCF (green, dotted line). The clustering algorithm assigns the grids to one of 15 clusters based on the similarity in the trend of active cell counts for each combustion application. The representative trend for each cluster is shown, while some random grids are shown faded in the background. The number of grids in each cluster ($N_A$, $N_B$, and $N_C$) is also shown.}
    \label{fig_reaction_clusters}
\end{figure}

Figure~\ref{fig_reaction_clusters} presents the clustered active cell count trends across the three combustion applications. We define $N_{\text{active}}$ as the number of cells that have not reached $t_{\text{final}}$ after $K$ integration steps. The figure displays representative trends from each cluster in bold colors, with 100 randomly selected grids from each cluster shown in faded colors. The number of grids in each cluster ($N_A$, $N_B$, and $N_C$ for the 14-species detonation, 30-species detonation, and JISCF cases, respectively) illustrates the relative frequency of different active cell decay patterns.

The results confirm the characteristic chemical stiffness patterns inherent to combustion simulations. For the JISCF case, clusters C1 through C6 exhibit steady decreases in $N_{\text{active}}$ with approximately 50\% of cells becoming inactive by $K=10$. This rapid deactivation reflects the highly variable chemical activity typical of turbulent reacting flows, where local mixing and temperature variations create regions of intense reaction followed by rapid completion.

The $14$-species detonation case exhibits markedly different behavior, with clusters $1-6$ displaying extended "long tails," where the majority of cells persist for hundreds to thousands of integration steps. This pattern reflects the more uniform chemical conditions characteristic of planar detonation structures, where reaction zones are spatially coherent and cells tend to undergo similar integration histories.

Notably, not all grids exhibit high chemical activity. Clusters around C10 and beyond show $N_{\text{active}}$ remaining nearly constant across all values of $K$, indicating grids where cells either fall below the minimum reaction temperature threshold or contain no reactive material. Cluster C15 represents grids with essentially no chemical activity across all three applications.

The steep decline in $N_{\text{active}}$ observed by $K=10$ for the most active clusters validates our choice of $K_{\text{max}}=5$ for the bulk integration phase in Algorithm~\ref{alg_optimized_integration}. However, the variable intervals between steep drops in different clusters suggest potential benefits from adaptive $K_{\text{max}}$ selection. Such an approach would require predictive algorithms capable of estimating completion times for individual cells—a topic for future investigation.

These patterns directly motivated the bulk-sparse integration strategy, as the majority of grids demonstrate significant chemical stiffness with highly non-uniform completion times. The dramatic variations in active cell counts across clusters confirm that traditional approaches, which integrate all cells uniformly, would suffer from substantial thread divergence and inefficient GPU utilization, validating the performance improvements demonstrated in Section~\ref{sec_results_multi_gpus}.

\section{Performance Comparison between cuBLAS and Naive Matrix Multiplication}\label{apx_naive_vs_cublas_dgemm}

\begin{figure}
    \centering
    \includegraphics[width=0.9\linewidth]{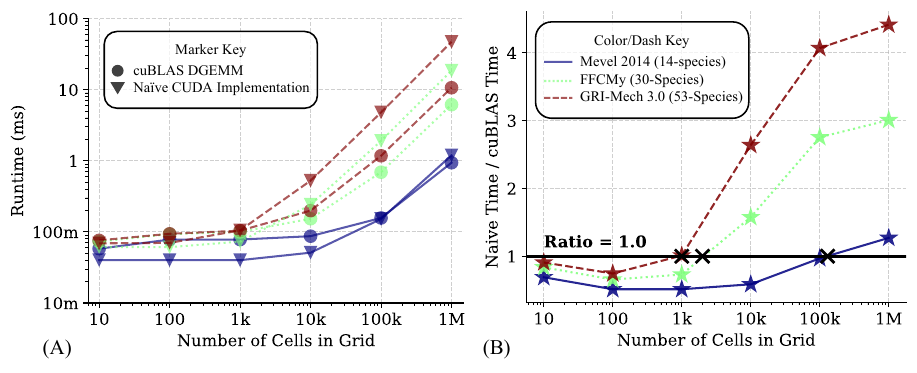}
    \caption{Performance comparison between cuBLAS DGEMM (circles) and naive CUDA matrix multiplication (triangles) for chemical kinetics operations across three mechanisms: 14-species Mevel (blue), 30-species FFCMy (green), and 53-species GRI-Mech (red). (A) Absolute runtime versus grid size. (B) Performance ratio showing cuBLAS advantage for larger grids. The black line indicates equal performance.}
    \label{fig_naive_vs_cublas_dgemm}
\end{figure}

To evaluate the tradeoffs in our single-kernel approach, as discussed in Section~\ref{sec_chemistry_algorithm_optimization}, we compared the performance of cuBLAS DGEMM with a naive CUDA matrix multiplication implementation for typical chemical kinetics operations. Figure~\ref{fig_naive_vs_cublas_dgemm} shows runtime comparisons across three chemical mechanisms with varying grid sizes representing typical AMR distributions.

For small grids (fewer than $10^3$ cells), the naive implementation performs comparably to cuBLAS due to insufficient work to saturate the GPU and the associated overhead of cuBLAS. However, as grid size increases beyond $\sim 10^4$ cells, cuBLAS demonstrates superior performance, achieving up to $4.5 \times$ faster execution for the largest grids tested. The crossover point occurs around $ 10^3$ to $10^5$ cells, depending on the complexity of the mechanism.

Figure~\ref{fig_naive_vs_cublas_dgemm} illustrates that cuBLAS advantages are most pronounced for larger mechanisms (53-species GRI-Mech), where the computational intensity justifies library overhead. However, our multigrid kernel fusion strategy must process hundreds of small grids simultaneously, where the naive approach's simplicity and reduced memory management overhead outweigh the benefits of cuBLAS optimizations. This analysis validates our design choice to prioritize kernel launch reduction over individual kernel efficiency in the AMR context.

\section{Performance Variation with Active Cell Count}\label{apx_performance_vs_active_cells}

\begin{figure}[H]
    \centering
    \includegraphics[width=0.7\textwidth]{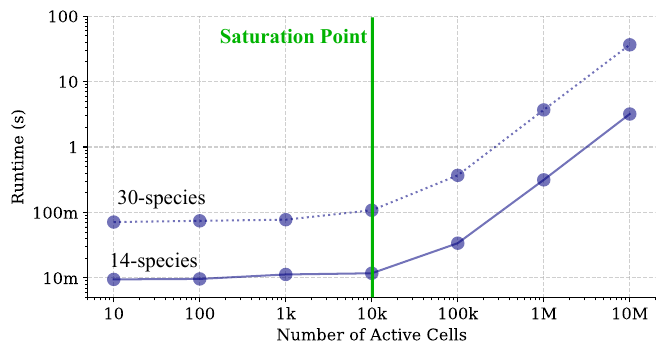}
    \caption{Single GPU runtime vs active cell count in one grid for the 14-species mechanism \citep{mevel2009hydrogen} and the 30-species mechanism \citep{smith2020foundational} for the isolated chemistry routine. The solid green vertical line indicates the point at which GPU saturation is reached.}
    \label{fig_runtime_vs_active_cells}
\end{figure}

The bulk integration phase of Algorithm~\ref{alg_optimized_integration} executes until the total active cell count across all grids falls below a threshold, $N_{\text{active}}^*$. This threshold was set to $10^4$, which is the point at which the number of GPU threads becomes saturated, as shown in Figure~\ref{fig_runtime_vs_active_cells}. $N_{\text{active}} = 10^4$ is a vital inflection point because grids with initial active cell counts greater than $10^4$ may see significant speedup. For instance, several grids in the JISCF case have an initial active cell count of $\sim 10^6$ (Clusters C$1$-C$5$ in Figure~\ref{fig_reaction_clusters}). By transitioning from a bulk to sparse integration strategy, we achieved a $\sim 500\times$ speedup in the chemistry routine for a single grid based on the $30$-species mechanism plotted in Figure~\ref{fig_runtime_vs_active_cells}.

\section{Griding Behavior and Chemistry Peformance Effects}\label{apx_grid_behavior}

\begin{figure}[H]
    \centering
    \includegraphics[width=\textwidth]{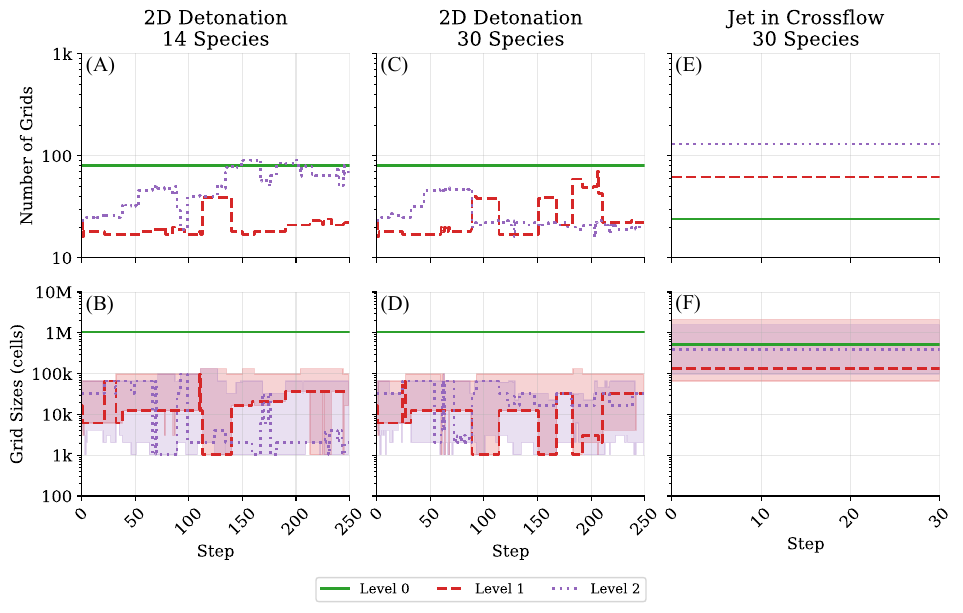}
    \caption{The number of grids and their size distribution as the simulation evolves for the 14-species 2D detonation (panels A and B), 30-species 2D detonation (panels C and D), and JISCF (panels E and F). Panels A/C/E show grid counts per AMR level. Panels B/D/F show, at each AMR level, the most common grid size, along with shaded envelopes for the smallest and largest grid sizes.}
    \label{fig_grid_size_distribution}
\end{figure}

The dynamic nature of AMR creates time-varying grid distributions that directly impact the performance of the chemistry routine. Figure~\ref{fig_grid_size_distribution} shows that the problems considered here generated $100-200$ total grids across all AMR levels, with individual grid sizes ranging from $10^3$ to $10^6$ cells. Note that regriding was disabled for the jet in supersonic crossflow case, resulting in a constant grid count and grid size distribution. This distribution pattern creates performance challenges for Algorithm~\ref{alg_naive_kinetics_solver}, where each grid launches separate chemistry kernels, leading to excessive kernel launch overhead and suboptimal GPU utilization when processing numerous small grids. To address these limitations, we implemented a single-kernel approach that processes all active cells across all grids simultaneously using the index mapping strategy from Algorithm~\ref{alg_optimized_integration}. 

\begin{figure}[H]
    \centering
    \includegraphics[width=\textwidth]{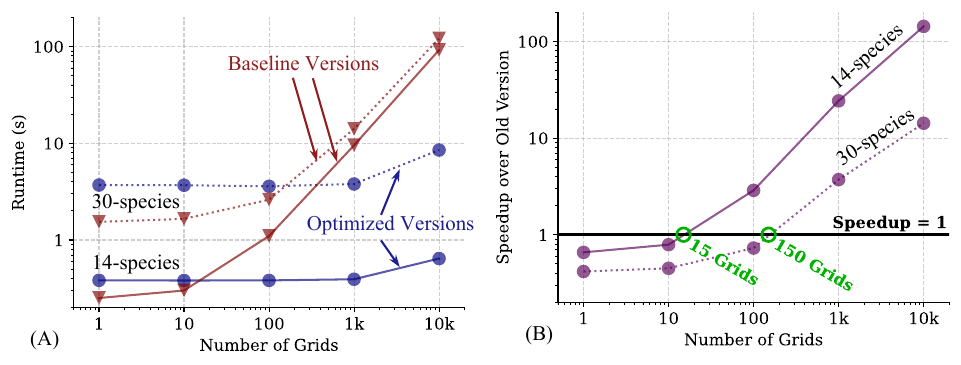}
    \caption{ (A) Kinetics runtime performance on 1 GPU with increasing number of grids at a constant total cell count of 1 million across all grids. Both the baseline and optimized versions of the code are shown. (B) Same data as (A), but the performance of the baseline version is divided by the performance of the optimized version to highlight the performance improvement. The green circles indicate the point at which the optimized version of the code becomes faster than the baseline version.}
    \label{fig_runtime_vs_n_grids}
\end{figure}

Figure~\ref{fig_runtime_vs_n_grids} demonstrates the effectiveness of this kernel fusion strategy, showing that performance improvements occur when grid counts exceed approximately 15 grids for the 14-species mechanism and 150 grids for the 30-species mechanism. Since most of our applications consistently operate above these thresholds, the optimized approach achieves up to $5\times$ speedup over the baseline implementation by eliminating kernel launch overhead and improving GPU resource utilization.

\section*{Acknowledgments}
This work was supported by the US Air Force Office of Scientific Research under grant number FA9550-23-1-0067 with Dr. Fariba Fahroo as Program Manager. IT support provided by the Lighthouse Advanced Research Computing team at the University of Michigan is greatly acknowledged. We thank Alireza Khadem for discussions on memory layout (~Section~\ref{sec_memory_optimizations}), which provided useful context in exploring different approaches. 

\bibliographystyle{elsarticle-harv} 
\bibliography{refs}

\end{document}